\documentclass[12pt]{iopart}

\usepackage{enumitem}
\usepackage{graphicx}%
\usepackage{tabularx}
\usepackage{multirow}
\usepackage{array}
\begin{document}

\title[Intercomparison of Open Electricity System Capacity Expansion Models]{Process and Policy Insights from an Intercomparison of Open Electricity System Capacity Expansion Models}

% \author{Content \& Services Team}

% \address{IOP Publishing, Temple Circus, Temple Way, Bristol BS1 6HG, UK}
% \ead{submissions@iop.org}
\author{%
Greg Schivley$^{1}$,
Aurora Barone$^{2}$,
Michael Blackhurst$^{3}$,
Patricia Hidalgo-Gonzalez$^{4}$,
Jesse Jenkins$^{1}$,
Oleg Lugovoy$^{5}$,
Qian Luo$^{1}$,
Michael J. Roberts$^{6}$,
Rangrang Zheng$^{6}$,
Cameron Wade$^{7}$,
Matthias Fripp$^{2}$
}

\address{$^{1}$Princeton University, Princeton, NJ, USA}
\address{$^{2}$Environmental Defense Fund}
\address{$^{3}$Carnegie Mellon University, Pittsburgh, PA, USA}
\address{$^{4}$University of California San Diego, La Jolla, CA, USA}
\address{$^{5}$Optimal Solution LLC, Richmond, VA, USA}
\address{$^{6}$University of Hawai'i, Honolulu, HI, USA}
\address{$^{7}$Sutubra Research Inc., Halifax, Nova Scotia, Canada}
\vspace{10pt}

\ead{greg.schivley@princeton.edu}

\begin{indented}
\item[]March 2025
\end{indented}

\begin{abstract}
This study performs a detailed intercomparison of four open-source electricity capacity expansion models—Temoa, Switch, GenX, and USENSYS—to evaluate 1) how closely the results of these models align when inputs and configurations are harmonized, and 2) the degree to which varying model configurations affect outputs. We harmonize the inputs to each model using PowerGenome and use clearly defined \emph{scenarios} (policy conditions) and \emph{configurations} (model setup choices). This allows us to isolate how differences in model structure affect policy outcomes and investment decisions. Our framework allows each model to be tested on identical assumptions for policy, technology costs, and operational constraints, allowing us to focus on differences that arise from inherent model structures. Key findings highlight that, when harmonized, models produce very similar capacity portfolios under current policies and net-zero scenarios, with less than 1\% difference in system costs for most configurations. This agreement among models allows us to focus on how configuration choices affect model results. For instance, configurations with unit commitment constraints or economic retirement yield different investments and system costs compared to simpler configurations. Our findings underscore the importance of aligning input data and transparently defining scenarios and configurations to provide robust policy insights.
% Through this study, we identify structural configurations that influence outcomes in all models. This work offers a valuable benchmark and identifies key modeling choices for policymakers, which ultimately will improve transparency and reliability in modeling efforts to inform planning for the clean energy transition.
\end{abstract}

%\maketitle

\section{Introduction}

Safely mitigating climate change requires deep decarbonization of energy systems in the next 15–30 years \cite{intergovernmental_panel_on_climate_change_ipcc_summary_2022}. The electricity sector will be key to this transition. To meet this challenge, stakeholders need tools to help plan for deep decarbonization of power systems on a national scale \cite{deep_decarbonization_pathways_project_pathways_2015}. Thirteen states, Puerto Rico, and Washington D.C. have statutes requiring 100\% carbon-free electricity supply within the next three decades \cite{clean_air_task_force_state_2020}, and investments in decarbonizing the US power sector will likely accelerate as a result of the Inflation Reduction Act passed in 2022. We are also in a time of rapid cost reductions for low-carbon power technologies due to continued R\&D advances and economies of scale, coupled with substantial uncertainty about grid technology and policy options. Better models can help planners and policymakers prioritize investments that support rapid and reliable decarbonization while minimizing costs and avoiding risks such as premature obsolescence. 

Power system planning has historically relied on proprietary models that were designed for an electricity system based largely on fossil fuels and are not well-suited for planning clean power systems. For example, wind turbines, solar panels, and batteries are declining in cost every year and are now able to provide power at a lower levelized cost than fossil-fueled generators, yet traditional capacity planning models based on load duration curves (probability distributions of load) cannot accurately model the simultaneous variation in time and space of variable resources (e.g., wind and sun) and load, or the inter-hour coupling introduced by energy storage or demand flexibility. 

A new generation of well-established open-source electricity system capacity expansion models (CEMs) has been created to aid both in designing power systems served by high levels of renewable energy and in modeling a diverse range of energy pathways. They are often used to study the impact of national-level policy decisions. Although these models may use many of the same underlying methods and constraints, differences in input data, model setup, or configurations--choices of model features, resolution, etc.--can lead to divergent findings. Unexplained differences in results from models that use similar approaches and methods can confound decision making. An improved understanding of why model results differ and whether these differences matter can advance the field and facilitate consensus when making policy decisions.

Both input and structural assumptions can lead to different model results. Commodity prices and infrastructure costs are examples of input assumptions that may vary by model but can be relatively easy to harmonize. The method of determining asset retirement, such as aging out assets at the end of their service life versus retiring assets based on economic uncompetitiveness, is a structural assumption. Methods to treat within-model input uncertainty and variability, such as sensitivity and Monte Carlo analysis, are both mature and common. These techniques examine how model results vary with differing input assumptions and, in doing so, can identify which input parameters are most influential or what results are robust to varying model inputs. In contrast, studies rarely perform experiments under a range of different structural configurations, partly because of the challenge of identifying varying structural methods from the perspective of a singular model.   

Intermodel comparisons can broaden the scope of structural and input modeling choices to encompass those made across different modeling teams. Previous intermodel comparison efforts, such as those by the Stanford Energy Modeling Forum \cite{huntington1982modeling} have been instrumental in developing and applying methods to fairly align policy scenarios assumptions across models. However, these efforts have not yet fully harmonized respective input assumptions or structural differences, which limits the ability to isolate the precise drivers of divergent model outcomes \cite{henry_promoting_2021, van_ouwerkerk_impacts_2022}. Henry et al. \cite{henry_promoting_2021} highlighted the often untested influence of structural differences--such as treatment of technology characteristics and optimization approaches--but only for a single simulation period, relatively simplified scenarios, and for only five greenfield generation technologies. Additionally, van Ouwerkerk et al. \cite{van_ouwerkerk_impacts_2022} demonstrated that structural features, such as the treatment of storage technologies or generator dispatch behavior, can lead to substantial discrepancies in model outcomes even under harmonized scenarios. Similarly, Sánchez-Pérez et al. \cite{sanchez-perez_effect_2022} depicts how the treatment of time drastically affects the optimal investment and operation of long-duration storage, and Jacobson et al. \cite{jacobsonQuantifyingImpactEnergy2024} demonstrates that renewable siting decisions can be suboptimal when using lower temporal resolution. Hoffmann et al. \cite{hoffmann_review_2024} emphasize that developing shared mathematical structures and consistent modeling frameworks can facilitate clearer communication and understanding between different models, promoting transparency and comparability. These studies suggest that not only does it remain difficult to fairly compare model results, but that there is not yet consensus on how to design and execute intercomparison studies. 

To address this gap, our study performs a rigorous comparison of four open-source, decarbonization-focused electricity capacity expansion models: Tools for Energy Model Optimization and Analysis (TEMOA)\cite{hunter_modeling_2013}, Switch\cite{johnston_switch_2019}, GenX\cite{jenkins_genx_2024}, and United States Energy SYStem (USENSYS)\cite{lugovoy_usensys_2024}. Unlike previous comparisons, we explicitly harmonize input data (using the PowerGenome data tool \cite{schivley_powergenomepowergenome_2024}). We systematically test different structural configurations across several relevant policy scenarios, and compare results across models and configurations within each of these scenarios. This approach enables us to clearly distinguish how structural modeling choices and assumptions influence model outcomes and investment decisions, and to identify residual differences attributable to the models’ internal formulations. It also allows us to identify which model features (as chosen in each configuration) are most important for studying different types of scenarios (e.g., current policies versus net-zero decarbonization). Our quantitative results enable us to draw highly specific and defensible conclusions about the sources of differences among modeling efforts. This approach allows us to provide clear and confident recommendations for policy decisions.

In addition to presenting quantitative model results we also use the intercomparison process to suggest guidelines and more generalizable methods for performing modeling intercomparisons. We conclude with a summary of key implications and recommendations for consumers of results/analysis from energy systems planning models and users of open-source CEMs.

The scenarios we model are of general interest and are not intended to be comprehensive. They are designed to test how different input assumptions and model configurations can impact the outcomes, and thus the policy recommendations. For instance, our net-zero scenario uses emission targets derived from a model that includes hydrogen production via electrolysis. However, we do not include the resulting electricity demand -- or any endogenous production of hydrogen -- in our systems. We do not explore strict zero-emission requirements without a buyout price, demand growth from data centers, the potential role of long-duration energy storage, alternative firm energy technologies, or the many other relevant policies, technologies, or interactions of the power system with the broader economy. These topics are not among the objectives of this effort. The main objective of this work is to perform an intercomparison of models with harmonized model inputs and explore the degree to which varying model configurations affect output.

% a set of sensitivities for scenarios with zero CO$_2$ emissions (not net-zero) and no carbon capture and sequestration is not explored. However, this set of scenarios is of high interest for policy makers and regulators as carbon capture and sequestration is not currently a commercially available and cost-competitive technology. Hence, it is of importance to understand which cost-competitive and commercially available technologies can play an enabling role in a zero emissions future. Although this is a pressing topic to delve into, it is not among the objectives of this effort as we do not intend to be comprehensive. The main objective of this work, as described before, is to understand how input assumptions and model configurations affect outcomes.

\section{Methods}
We harmonized and compared four energy system models representing the electric power system for the continental US: GenX, Switch, USENSYS, and TEMOA. All four models employ optimization to select the least cost set of energy sources and technologies for energy conversion, transmission, storage, and carbon management that meet electricity demand on an hourly basis, subject to exogenously specified costs, policies, and scenario assumptions. All of them seek to minimize expenditures during the study period, assuming capital costs are amortized over the life of the asset (not paid immediately); when run in multi-period foresight mode, they seek to minimize the discounted total (net present value) of these costs across all years of the study.

The study relies on the concepts of \emph{scenarios} (the world we optimize for) and \emph{configurations} (choices of time and space resolution, technology availability, and model features to use when setting up the model). The scenarios are used to represent policy questions of interest. In general, users get an answer to a policy question by comparing scenarios to each other, e.g., comparing emissions and costs between the deep-decarbonization scenario and the current policies scenario. 

Configurations allow us to distinguish between the effect of choices users make when setting up their models and differences in how features are implemented in the models themselves. In other comparison studies, much of what is perceived as “model differences” are actually due to different configuration choices made by different modeling teams. The models might produce much more similar results if they were run using the same configurations. We avoid this ambiguity by using the same configurations for all models. This allows us to identify which specific features are important for effectively modeling each scenario, and also to assess the extent to which internal differences between the models drive differences in results. 

Models were harmonized in three stages: (1) prospectively aligning assumptions and modeling approach prior to optimization; (2) revising models by comparing results for a common base-case scenario and base-case configuration; and (3) revising models by comparing results across a broader set of scenarios and configurations. 

\textbf{Prospectively aligning model assumptions:} We used the output specifications of PowerGenome \cite{schivley_powergenomepowergenome_2024} as a guide to align model assumptions prior to any simulations. PowerGenome is a data management system originally developed to provide inputs for GenX. Using the same source data significantly narrows any differences between models. \Tref{table1} and \Tref{table2} summarize the common structural and input assumptions made, respectively, prior to any simulations. In addition to the assumptions listed in \Tref{table2}, we clustered fossil-based generators within each region based on their heat rate and fixed operating costs. Clustering reduces the dimensionality of models while preserving the underlying variability in resources. 
% We then averaged model inputs by region, planning horizon, and either cluster (for thermoelectric fossil generators) or technology (for all other generators).

\begin{table}[!ht]
\caption{\label{table1}Primary structural assumptions used to harmonize models.}
    % \begin{indented}
    \footnotesize
    \begin{tabular}{@{}p{1.5in}p{4in}}
    \br
        \textbf{Feature} & \textbf{Structural assumption} \\ \hline
        Spatial scope & Continental United States \\ 
        Spatial resolution & 26 zones as depicted in \Fref{fig:model-regions} \\ 
        Planning horizon & 2024 through 2050 \\ 
        Planning periods & Initial periods in 2024-27 and 2027-30, then 5-year increments through 2050 (6 periods in total). \\ 
        Representative periods & 52 independent weeks per year (myopic across years) or 20 independent weeks (foresight across years) \\ 
        New-build technologies & 
        \begin{minipage}[t]{\linewidth}
        \begin{itemize}[nosep,after=\strut]
            \item Li-Ion batteries 
            \item Hydrogen combustion turbine 
            \item Hydrogen combined cycle
            \item Natural gas combustion turbine 
            \item Natural gas combined cycle (NGCC) 
            \item Natural gas combined cycle with 95\% CO$_2$ capture (CCS) 
            \item Conventional nuclear 
            \item Onshore wind 
            \item Offshore wind 
            \item Utility-scale solar
        \end{itemize} 
        \end{minipage}\\ 
        Power system operations & Peak demands are not met when generation costs exceed \$5000/MWh.\newline
        Unit commitment operational constraints are only included in specific scenarios.\newline
        No resource adequacy requirements are included beyond the \$5,000/MWh unserved load penalty. \\ 
        Subsidies & Investment tax credits are applied to the capital cost or annuity, production and carbon sequestration tax credits are applied to the variable cost per MWh \\
        \br
    \end{tabular}
    % \end{indented}
\end{table}
\normalsize

\begin{table}[!ht]
\caption{\label{table2}Primary input assumptions used to harmonize models}
    % \begin{indented}
    \footnotesize
    \begin{tabular}{@{}p{1.5in}p{4in}}
    \br
        \textbf{Parameter set} & \textbf{Empirical Basis} \\ \hline
        Fuel prices & EIA AEO \cite{us_energy_information_administration_annual_2023} is used for most prices. \$16/MMBTU is assumed for hydrogen. \\  
        Existing generators & Capacity and heat rates from PUDL \cite{selvans_pudl_2022}, using EIA 2022 annual data and EIA 860m from June 2023 \cite{us_energy_information_administration_preliminary_nodate_860m} \\ 
        Capacity factors & Wind and solar hourly profiles are from Vibrant Clean Energy \cite{vibrantcleanenergyllcUSWindSolar2024}. Hydro inflow is calculated as a rolling average of monthly production in 2012. \\ 
        Load profiles & Historical profiles from NREL \cite{mai_electrification_2018} projected using EIA AEO sector growth rates \cite{us_energy_information_administration_annual_2023} and Princeton’s REPEAT \cite{jenkins_climate_2023} to account for increased electrification of end-use technologies \\ 
        Capital costs & NREL ATB \cite{national_renewable_energy_laboratory_annual_nodate} is used for most equipment costs, except we use a WACC of 5\% for all technologies. Regional costs for each technology are adjusted using factors from EIA \cite{us_energy_information_administration_cost_2020}. Hydrogen generators are assumed to cost 10\% more than natural gas generators \cite{christidis2023h2}. Interconnection costs for wind and solar are estimated using methods described in \cite{patankar_land_2023}. CO$_2$ pipeline capital costs are estimated using NETL \cite{national_energy_technology_laboratory_fecmnetl_2022}. \\ 
        Discount rate & Foresight models assume a 2\% discount rate. \\ 
        Subsidy amounts & Investment and production tax credits from the IRA in current policy scenarios \cite{internal_revenue_service_inflation_nodate}. Production tax credits are adjusted to an annualized value over the full lifetime of the facility. A full description is provided in \ref{appendix:PTC}. \\
        \br
    \end{tabular}
    % \end{indented}
\end{table}
\normalsize

\textbf{Harmonizing models for the net-zero scenario and base configuration:} We used an iterative and exploratory process to identify and resolve differences in model results for the net-zero scenario with the base configuration, which are summarized in Tables 3 and 4. We used the following results to compare models: generation and transmission capacity, emissions, generation by energy sources, and total system cost in an operational simulation. Given the size of the system and the number of possible substitutional investments, some differences in capacity decisions are expected. System costs from the operational simulation were used to confirm that the models solved the same problem to within the desired level of tolerance. Harmonization was considered sufficient when any residual differences in results--specifically, the energy mix and geographical disposition of infrastructure builds and total costs--did not show large, unexplainable differences across models. In most cases, differences could be traced to errors in input data, unintended configuration differences, or subtle differences in the way investment windows and plant retirements were characterized. Thus, intercomparison helped to flag errors and improve harmonization of intended configurations and scenarios.

The net-zero scenario examines what is required to follow the most direct feasible path to achieving near-zero carbon in the electricity sector in 2050 and the base configuration includes a subset of features common to all models. Although the decarbonization pathway in this base scenario is aggressive, following the REPEAT study \cite{jenkins_climate_2023}, it includes a buyout price of \$200 per ton for emissions above the targets to account for alternative mitigation strategies. We employed an open-ended, consensus-based approach to make specific changes to individual models. The primary changes made during this stage included configuring all models to:
\begin{itemize}
    \item Exclude any capacity associated with generators that retire within a planning horizon;
    \item Consistently represent constraints on reservoir hydroelectric generation;
    % \item Consistently represent state and regional renewable portfolio and clean energy standards; 
    \item Use the same cost of capital (WACC) for all assets (generators, transmission lines, pipelines, etc.);
    \item Use same asset life for generator interconnections and CO$_2$ pipelines as for the generator itself; 
    \item Do not derate power plants based on planned or forced outage rates;
    \item Use the same optimization solver settings (Gurobi barrier method with no crossover); 
    \item Allow unserved load with a \$5000/MWh penalty; and
    \item Allow nuclear and coal plants to ramp up and down like other plants (not must-run or fixed-output).
\end{itemize}

\textbf{Harmonizing models for additional scenarios and configurations:} Tables 3 and 4 respectively list all the scenarios and configurations modeled. Select, but not all, combinations of scenarios and configurations were modeled due to the intensive computing requirements for each case. Comparing the effectiveness (i.e., final cost) of each configuration in each scenario is intended to show which features are important for modeling each scenario. In the original project design, not all models were expected to be able to implement every configuration, so the study would give some insight into which models were best suited for which scenarios. However, over the course of the project, most or all of the models were extended and adapted as needed to include multi-period modeling with foresight, myopic multi-period modeling, and economic retirement. The models were also adjusted to give similar treatments of carbon targets (with buyout), clean energy standards, minimum-capacity targets, limits on transmission expansion, CCS subsidies, production tax credits, and investment tax credits. In some cases, this required careful attention to data that were provided in different units to different models--such as hydro flow limits or capital costs (overnight cost versus annuity)--to ensure the values were equivalent across models.

\begin{table}[!ht]
    \caption{\label{table3}Summary of modeled scenarios}
    \footnotesize
    \begin{tabular}{@{}p{0.75in}p{2.75in}p{2.5in}}
    \br
    \textbf{Parent Scenario} & \textbf{Description} & \textbf{Child scenarios} \\
    \hline % Thin line under the column headers
    Current policies & Existing national and state policies. Policies include investment tax credits for solar, production tax credits for zero-carbon generation (see \ref{appendix:PTC} for details), subsidies for hydrogen and CCS, local CO$_2$ caps, renewable portfolio/clean energy standards (RPS/CES), and minimum capacity targets for offshore wind. & None \\
    Net-zero & National CO$_2$ cap on net-zero trajectory by 2050 (described in \ref{appendix:co2_targets}) with \$200/ton buyout price. No constraints on transmission and no other emission or clean energy policies. & 
    \begin{minipage}[t]{2.5in}
        1) Base (no changes)\\[0.5em]
        2) Carbon buyout prices of \$50 per ton or \$1,000 per ton\\[0.5em]
        3) Transmission expansion constraints of 0\% (none), 15\% (moderate) and 50\% (high) per corridor per period or 400 MW, whichever is greater\\[0.5em]
        4) No CCS allowed
    \end{minipage} \\
    \br
    \end{tabular}
\end{table}

\begin{table}[!ht]
\caption{\label{table4}Summary of modeled configurations}
    % \begin{indented}
    \footnotesize
    \begin{tabular}{@{}p{2in}p{3.75in}}
    \br
    \textbf{Configuration} & \textbf{Alternative specifications} \\ \hline
        Base & No requirement for unit commitment or limits on power plant ramp rates, minimum load, etc. Plants retire when they reach standard ages. We  optimize each study period independently, in succession, without consideration of later periods (myopic). We consider 52 independent weeks, spanning a full year of weather data. Model periods are 2024-27, 2028-30, and then every 5 years through 2050. Investment decisions are made to meet demand in the final year of each model period.\\ 
        Unit Commitment and Ramping & Base configuration using linearized unit commitment; including restrictions on ramp rate, min load, up and down time, and startup costs. \\ 
        Economic Retirement & Base configuration without age-based retirement of any existing plants; instead we retire generators permanently if the operating costs are higher than the revenue; avoidable costs include fixed O\&M but not capital recovery; capital recovery is amortized over normal life, but then continues through the whole study at the same rate. \\ 
        Short Sample Period & Base configuration using 20 weeks sampled from full weather period (with probability weights) instead of 52 weeks to reduce the model size. \\ 
        Multi-Period Foresight & Short Sample Period configuration where we co-optimize all 6 model periods together (short sample is needed to reduce allow for multi-period). \\ 
        Operational Simulation & Specialized case used to evaluate the performance and system cost of proposed plans in a common context. Base configuration, but freeze generation \& transmission decisions based on a previously run capacity expansion model, use limits from the Unit Commitment and Ramping configuration, and consider all 52 weeks of weather. Then run the model to simulate performance of the plan through the year. Any generation shortfalls are priced at \$5,000/MWh. \\
        \br
    \end{tabular}
    % \end{indented}
\end{table}

\textbf{Calculating model costs:} One purpose of this study is to show which configurations are most effective in identifying low-cost solutions in each scenario. Differences in how each model calculates and reports costs make it difficult to directly compare using outputs from each model. To address this difficulty, we run an operational simulation using a single model (GenX) with fixed capacity decisions, a full year of weather data, and constraints on cycling and ramp rates to calculate the cost of adopting the portfolio proposed by each model. Investment annuities are included as part of fixed operations and maintenance. (This is the “Operational Simulation” configuration in \Tref{table4}.) Comparing the total system costs of capacity decisions from each model serves as another harmonization check by revealing instances where different capacity decisions are substitutions that lead to the same system cost.

\section{Results}\label{section:results}

We use the harmonized models to compare within- and across-model differences for different scenarios and configurations. We profile detailed results for only a select set of comparisons chosen for policy relevance and to highlight the value of model intercomparisons. 

\subsection{Demonstration of Harmonization}

\begin{figure*}[ht]
\includegraphics[scale=1,trim={0cm 0cm 0 0cm},clip]{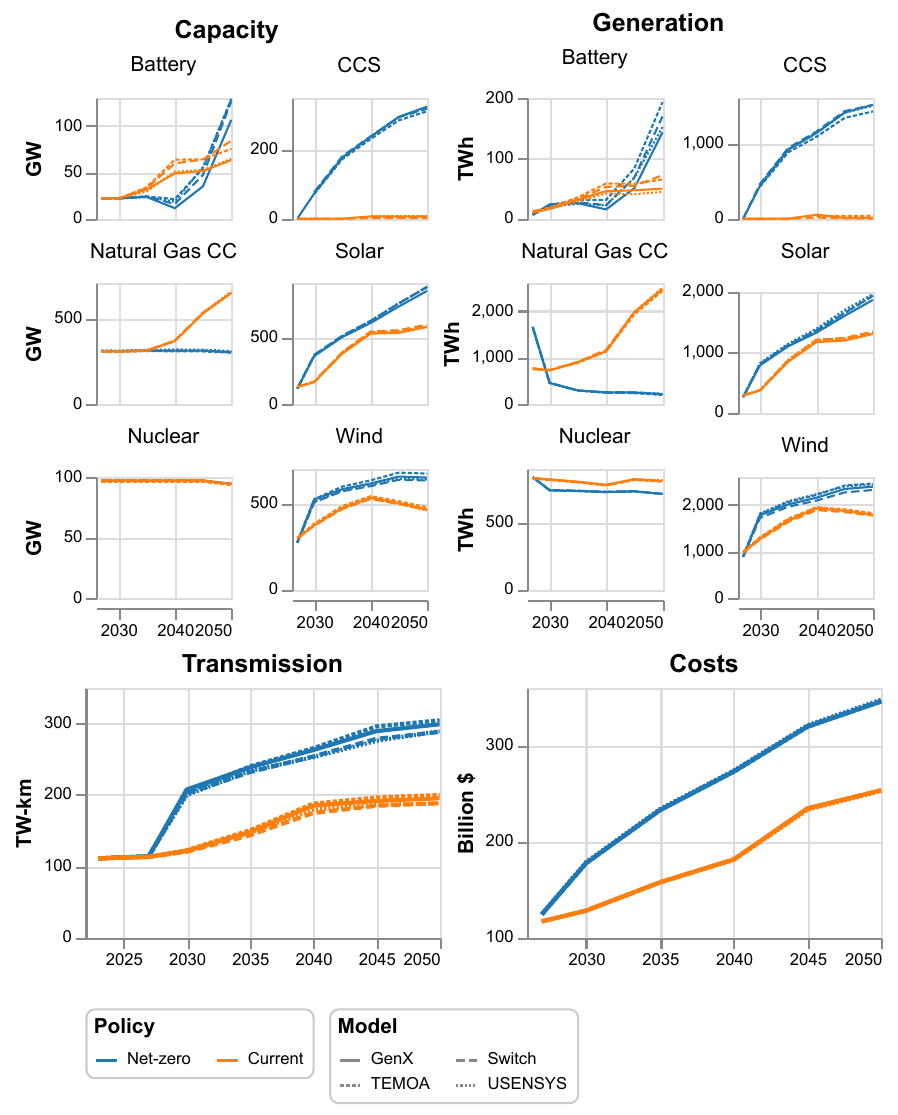}
\caption{ Results from each model under the base configuration of Net-zero and Current Policy scenarios. Subplots show total capacity (top-left) and generation (top-right) of selected resources, total transmission capacity (bottom-left), and annual operational system cost (bottom-right).}
\label{fig1}
\end{figure*}

We first demonstrate the effectiveness of our harmonization approach by comparing outcomes from the models under fully harmonized inputs and base-case assumptions. The net present values of the operational model system costs for the current policy and net-zero scenarios vary by 0.2-0.3\% across the models, demonstrating that models are all finding nearly equivalent global cost minimums. \Fref{fig1} summarizes across-model differences in generator capacities, generation mix, and optimized transmission capacities associated with the current and net-zero scenarios. Note that we refer to natural gas combined cycle turbines with 95\% CO$_2$ capture as “CCS” throughout. Power system optimization models can find different solutions with similar costs.  From this cost similarity, we conclude that the observed differences reflect normal quasi-random variation due to marginal differences in the inputs, algorithms or solution path used by each model. Model users should take note that variations on this scale are possible when using any model; they do not reflect persistent biases on the part of individual models, which each sought only to minimize costs.

\subsection{Comparing Model Scenarios}

\begin{figure*}[ht]
\includegraphics[scale=1,trim={0cm 0cm 0 0cm},clip]{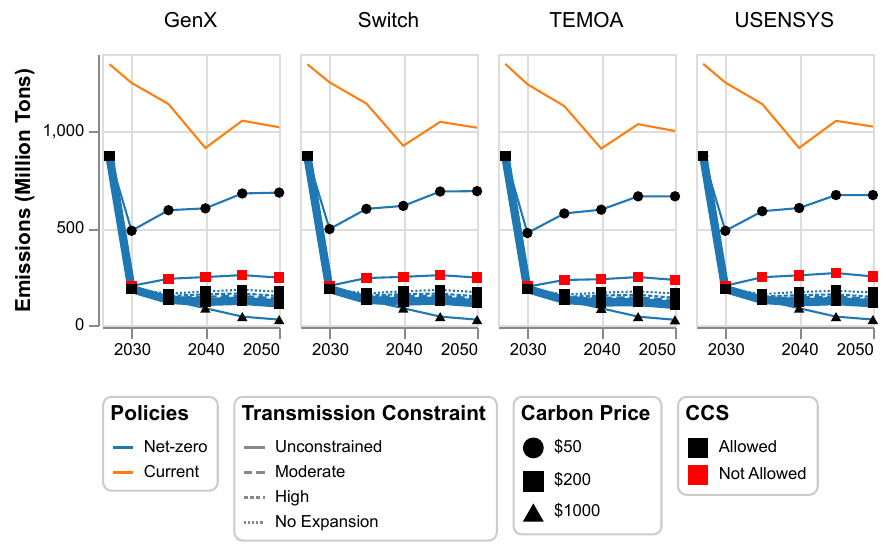}
\caption{Projected annual emissions within each planning period for the current policies and net-zero emissions scenarios. The color of the lines indicates current policies versus net-zero cases; the shape of the marker indicates CO$_2$ buyout price; dashed lines indicate limits on transmission expansion; and marker color indicates whether CCS is allowed. The base net-zero configuration is indicated by a thicker blue line. }
\label{fig2}
\end{figure*}

Next, we examine how model results differ across various policy scenarios, illustrating the impacts of varying scenario assumptions such as carbon buyout prices, transmission constraints, and technology availability. Note that there are several child scenarios derived from the net-zero parent scenario as summarized in Table 3. Across-model differences under those child scenarios are small, demonstrating that harmonized models converge to similar results. \Fref{fig2} shows emissions across models by scenario. Our results suggest that current policies are insufficient to achieve deep and sustained emission reductions, similar to previous studies \cite{bistline_emissions_2023, bistline_power_2023}.

\subsubsection{Steady emissions reductions would require a carbon buyout price of at least  \$1,000/tonne }

Among the tested net-zero child scenarios, the carbon buyout price has the most substantial impact on system emissions. The net-zero scenarios in this study have an exogenously specified emissions cap with a steep drop from 2027 to 2030. This trajectory is based on results from the net-zero scenario in Princeton’s REPEAT study, which found that the most cost-effective route to net-zero CO$_2$ emissions by 2050 required the electricity sector to reduce emissions earlier and faster than the rest of the economy \cite{jenkins_climate_2023}. In the base case, we allowed buyout from this cap at a cost of \$200 per tonne of CO$_2$. This is close to the current EPA estimate of the social cost of carbon emissions \cite{us_environmental_protection_agency_report_2023} and also the consensus view of the future cost of direct air capture of CO$_2$ \cite{howard_consensus_2024}. Variation in the assumed carbon buyout price has the most significant impact on the net-zero emissions scenario. A buyout price of \$200 per ton results in slowly decreasing emissions after 2030. Even by 2050, none of the models find emissions dropping below 100 million tonnes per year. At this buyout price nearly all capacity growth is in wind, solar, batteries, and CCS. A buyout price of \$50 per ton builds slightly less wind and solar capacity. The models generally favor expansion of conventional natural gas generation in place of CCS, explaining the steady increase in emissions after 2030. In contrast, a price of \$1,000 per ton favors expansions of CCS, wind, solar, and hydrogen, yielding steady emission reductions despite strong load growth. 

\subsubsection{Constraining transmission expansion increases emissions and costs due to instead paying a penalty per ton of CO$_2$ emitted
}

\begin{figure*}[ht]
\includegraphics[scale=1,trim={0cm 0cm 0 0cm},clip]{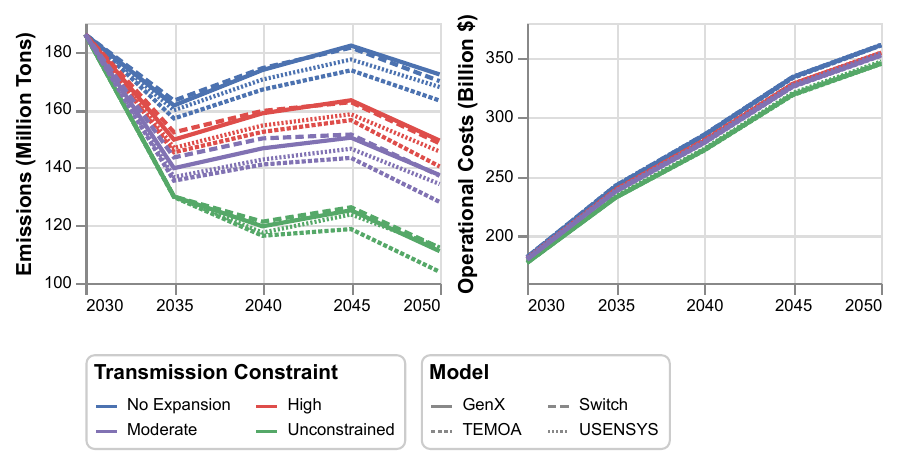}
\caption{Tighter constraints on transmission expansion increase both system costs and emissions. Results are identical in all constraint child scenarios prior to 2030. Costs and emissions are annual values for each planning period.}
\label{fig3}
\end{figure*}

The ability to build inter-regional transmission represents a policy choice; we find that constraining transmission expansion increases emissions and costs, as shown in \Fref{fig3}. Specifically, preventing all transmission expansion increases 2050 operational costs by roughly 4.6\% and increases emissions (approximately 55\%) due to reliance on fossil-based generation. \Fref{fig3} shows that constraining transmission expansion increases both emissions and costs. Relative to unconstrained expansion, preventing expansion increases operational costs in 2050 by 4.6\% and increases emissions by 55\%. This implies that it is possible to substitute for transmission by instead paying the ‘unmet policy penalty’ (cost per ton of CO$_2$ emitted). However, the total penalty appears to be modest, as is the overall cost of constraining transmission to suboptimal expansion. The most substantial substitute for transmission is battery capacity, but generation mixes and quantities also change within regions. \Fref{fig3} also shows more variation across transmission scenarios than between models, which partly speaks to the high substitutability of resources in meeting stringent decarbonization objectives. See Zheng et al. \cite{zheng_optimal_2024} for further discussion of these issues. 

\begin{figure*}[ht]
\includegraphics[scale=1,trim={0cm 0cm 0 0cm},clip]{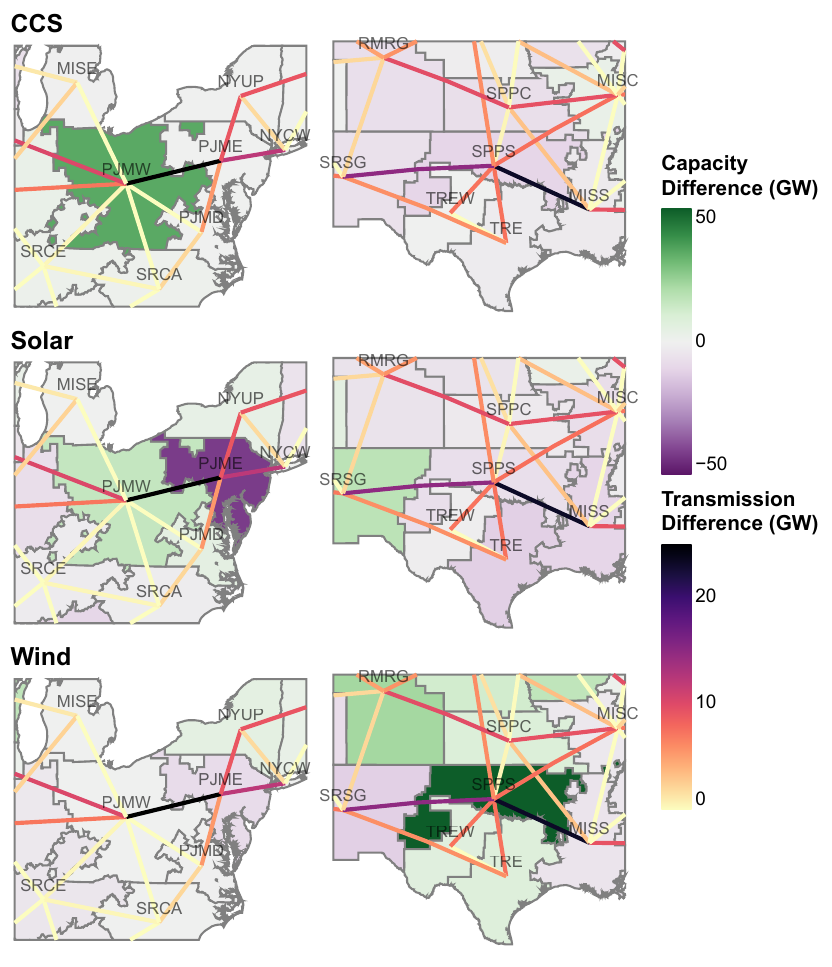}
\caption{Trade-offs between expansions of transmission and capacity of CCS, solar, and wind resources in select modeled regions as estimated by GenX for 2050. Results reflect  differences in resource capacity and transmission capacity between unconstrained and no (0\%) transmission expansion. Transmission constraints increase resource capacity in the purple regions and decrease it in the green regions.}
\label{fig4}
\end{figure*}

\Fref{fig4} shows the resource capacity substitutions that models make when no additional transmission can be built. Note that there are only a few key lines where the constraint has a major effect on aggregate transmission capacity in 2050: PJM - East (PJME) to PJM - West (PJMW) in the Northeast, and Southwest Power Pool - South (SPPS) to Midcontinent ISO - South (MISS) and SPPS to WECC - Southwest (SRSG) in the South. Without transmission constraints, the models prefer to build additional CCS and solar capacity in PJMW. When they cannot expand transmission between PJMW and PJME, they build more solar capacity in PJME. Unconstrained transmission favors the expansion of wind capacity in SPPS. When transmission out of SPPS is constrained, the model substitutes modest amounts of wind, solar, and CCS in adjacent regions.

\subsubsection{CCS as a technology option in a deep decarbonization scenario when no other cost-effective firm generation source is available}

\begin{figure*}[ht]
\includegraphics[scale=1,trim={0cm 0cm 0 0cm},clip]{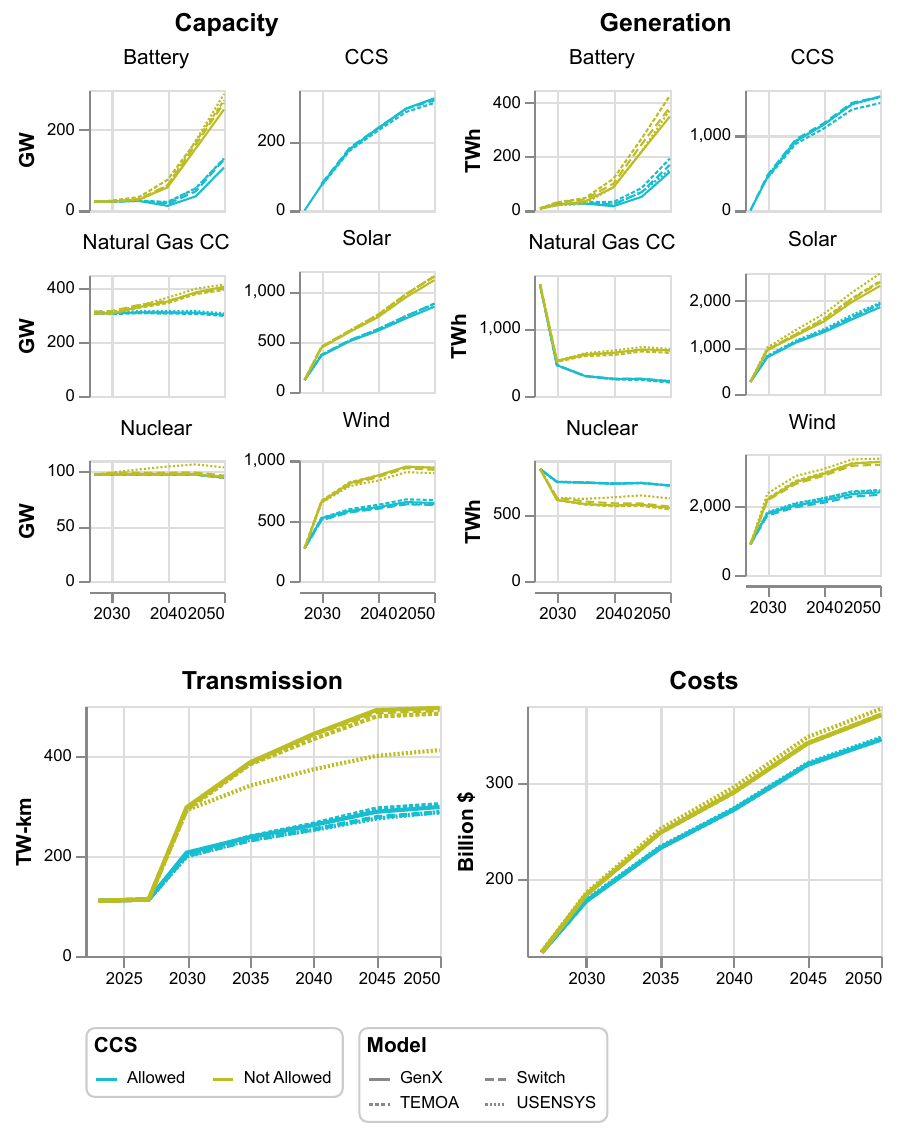}
\caption{Effects of allowing or disallowing CCS on total capacity, generation mix, emissions, transmission expansion, and costs under the Net-zero scenario. Capacity and transmission represent existing stock at the end of each planning period. Generation and costs are annual values within each planning period.}
\label{fig5}
\end{figure*}

Finally, we consider the scenario where carbon capture and storage (CCS) is unavailable, testing its role as a key firm generation source under deep decarbonization constraints. \Fref{fig5} shows the impact of removing CCS as a technology option in a deep decarbonization scenario when no other cost-effective firm generation source is available. All models show similar results: if CCS is not allowed, the ~300 GW of CCS in the base scenario would be replaced by 150 GW more battery, 300 GW more wind, 250 GW more solar, 100 GW more NGCC, and 35 - 58\% more transmission capacity. This results in about 134 million tons more emissions in 2050, 122\% higher than when CCS is available. Notice that these results hold in the context of: modeling CCS as technology with characteristics that are not currently commercially available, allowing a buyout of CO$_2$ emissions (as opposed to capping emissions to zero), and not considering other cost-effective firm generation sources. Disallowing CCS also raises the NPV social buyout cost of CO$_2$ emissions by around 200 percent (\$378 billion) relative to the base case. About 20\% of this extra CO$_2$ cost is offset in the form of lower total expenditures on fuel and equipment.

\subsection{Comparing Model Configurations}

\begin{figure*}[ht]
\includegraphics[scale=1,trim={0cm 0cm 0 0cm},clip]{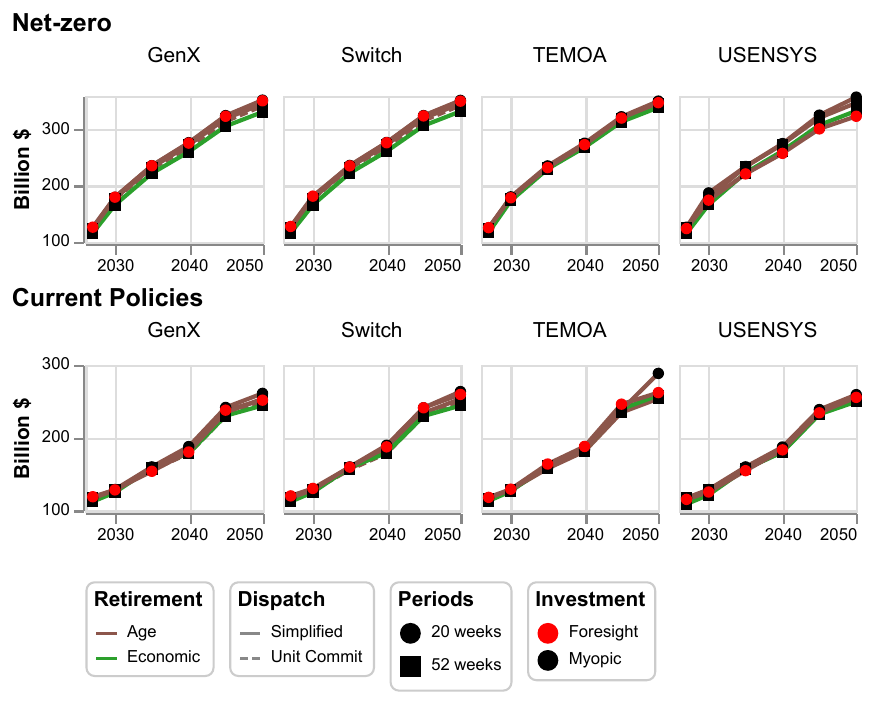}
\caption{Annual expenditure per planning period in 52-week operational simulation for each model configuration, assuming net-zero emissions (top panel) and current policies (bottom panel). The base configuration uses a solid brown line with square, black markers, indicating age-based retirement, simplified dispatch, 52 sample weeks and myopic outlook. Alternative configurations test deviations from each of these.}
\label{fig6}
\end{figure*}

Having explored the differences across scenarios, we now examine how configuration choices—such as retirement policies, temporal sampling, and foresight—affect model outcomes. \Fref{fig6} shows differences in cost for the model configurations we tested. Not all configurations were coded in all models. The models again produce relatively consistent results with some exceptions. 

Using economic retirement instead of age-based retirement is the most important single feature that we tested that allows models to find a solution with lower operational costs in both the current policies and deep decarbonization scenarios. Including unit commitment, ramp rate limits and minimum up- and down-time for generators in the capacity planning stage are also helpful for finding low-cost solutions. 

Running models with foresight across all investment periods could in principle lead to lower-cost solutions, as the models anticipate future changes in policies or costs for fuel and equipment. However, in practice, the size of large-scale electricity expansion models such as we used in this study is limited by the memory on available computers. Foresight models, which represent multiple study periods at one time, are larger than the individual stages of myopic ones, so running them typically requires using a smaller number of sample weeks. This failure to consider the full range of weather conditions can reduce the optimality of the solution. For this project, we used 20 sample weeks for the foresight models instead of the full 52 weeks of available data used in the base configuration. \Fref{fig6} shows that using 20 sample weeks in myopic mode (black circle markers) results in costs about 3\% higher than the base case. Switching from myopic to foresight mode for the 20-week cases improved the cost of the final plan slightly (red circle markers), but was generally not an improvement on the 52-week myopic base case (black square markers).

Although different models and configurations produced substantially similar cost capacity expansion plans within each scenario (generally within about 10\% across all configurations and models), the details of these plans and their emission levels can vary quite strongly. This reflects an overall principle observed in this work that a range of solutions can all have similar costs, so users should not assume that the details of any particular plan are set in stone. Different choices of model, configuration or even small differences in input data can result in quite different plans, even if they are equally good, as measured by the optimality of costs.

\begin{figure*}[ht]
\includegraphics[scale=1,trim={0cm 0cm 0 0cm},clip]{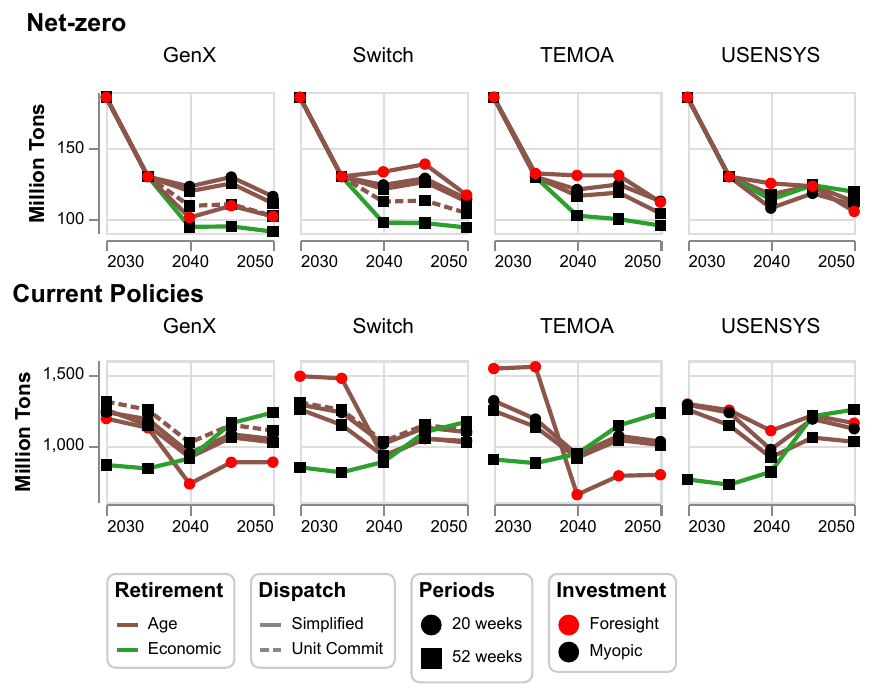}
\caption{Projected annual emissions per planning period for modeled configurations assuming net-zero emissions (top panel) and current policies (bottom panel). The base configuration uses a solid brown line with square, black markers, indicating age-based retirement, simplified dispatch, 52 sample weeks and myopic outlook. Alternative configurations test deviations from each of these.}
\label{fig7}
\end{figure*}

\Fref{fig7} shows emissions in each year for each model, for each configuration choice tested. The treatment of retirement consistently impacts emissions across models but in ways that differ by scenario. In the current policy scenario, economic retirement initially reduces emissions by retiring more than half of existing coal capacity. Beginning in 2040, after the phase out of existing policies, the economic retirement of nuclear plants, coupled with retained coal capacity, results in higher emissions relative to the base configuration. In the net-zero scenario, economic retirement leads to a near immediate elimination of coal generation and approximately 30\% reduction in natural gas generation, which explains much of the impact on emissions. The lost fossil generation is complemented by an increase in battery use and preservation of hydroelectric capacity that otherwise would retire due to age. Additional findings from different configurations are included in Table 5.

\begin{figure*}[ht]
\includegraphics[scale=1,trim={0cm 0cm 0 0cm},clip]{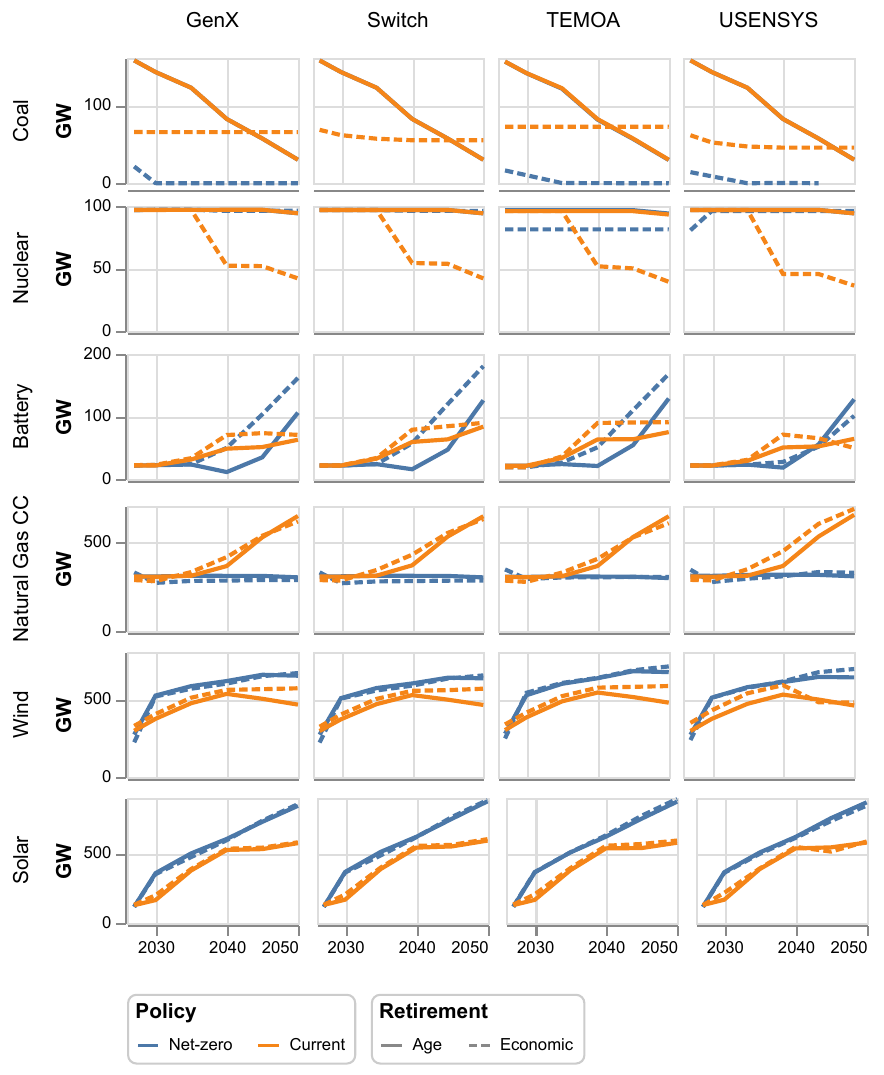}
\caption{Capacities at the end of each planning period in net-zero and current policy scenarios assuming age-based or economic retirement of existing generators.}
\label{fig8}
\end{figure*}

\Fref{fig8} shows how retirement configurations affect projected capacity across models. Assuming generators retire economically leads to an immediate retirement of about 50\% and 85\% of existing coal capacity for the current policies and net-zero scenarios, respectively. Economic retirement of coal capacity in the current policies scenario reduces its generation by approximately 500 TWh in 2027, most of which is replaced by increases in natural gas and wind generation. While the affected generators otherwise remain serviceable, the models find that it is more economically efficient to reduce total capacity and run the remaining plants at higher capacity factors. Similarly, 40\% of existing nuclear capacity is economically retired in the planning period 2040 after their subsidies expire. The subsequent reduction in nuclear generation compared to the base configuration in 2040 (340 TWh) is accompanied by a 90 TWh increase in CCS generation and 70 TWh of additional wind generation. Coal generation stays flat from 2035-2040 under economic retirement – and with AEO reference fuel prices – but would see a drop of 290 TWh in the base configuration. Economic retirement has a complicated effect on emissions in the current policies scenario – early coal retirements lead to emission reductions in early years, but investments to maintain coal plants and the retirement of nuclear capacity in 2040 lead to higher emissions in 2045-50. We discuss the relationship between fuel prices and current policy retirements in \ref{appendix:fuel_prices}. Effects on emissions are more straightforward in the base scenario, where allowing less efficient generators to retire early leads to lower emissions from 2040-2050.

\begin{table}[!ht]
\caption{\label{table5}Summary of modeled configurations}
\footnotesize
\begin{tabular}{@{}m{0.9in}m{2.4in}m{2.75in}} % All columns use "m{}" for consistent alignment
\br
\textbf{Configuration} & \textbf{Net-Zero} & \textbf{Current Policies} \\
\br
\raggedright \vspace{0pt} Unit Commitment and Ramping & 
\begin{minipage}[t]{\linewidth}
    \vspace{-2em} % Pull content up to align with top of the row
    \begin{itemize}[leftmargin=*, noitemsep, topsep=0pt]
        \item 8–12\% lower emissions per period starting in 2040.
        \item More new wind and battery capacity.
        \item Less new solar (all periods) and CCS (after 2040) capacity.
    \end{itemize}
\end{minipage} & 
\begin{minipage}[t]{\linewidth}
    \vspace{-2em}
    \begin{itemize}[leftmargin=*, noitemsep, topsep=0pt]
        \item Higher emissions in all periods (4\% in 2030, growing to 10\% in 2050).
        \item More new battery capacity.
        \item Less solar and NGCC capacity.
        \item More NGCC generation.
    \end{itemize}
    \vspace{0.1em}
\end{minipage} \\ 
\raggedright Economic Retirement & 
\begin{minipage}[t]{\linewidth}
    \vspace{-\baselineskip}
    \begin{itemize}[leftmargin=*, noitemsep, topsep=0pt]
        \item Lowers emissions slightly (2040 and later).
        \item Old fossil capacity is allowed to retire when CO$_2$ constraints limit their generation.
        \item In the hours where it is most cost-effective to meet demand with thermal capacity, newer and more efficient units can be used.
        \item Most existing coal immediately retires when allowed.
        \item Slightly less transmission built.
    \end{itemize}
\end{minipage} & 
\begin{minipage}[t]{\linewidth}
    \vspace{-\baselineskip}
    \begin{itemize}[leftmargin=*, noitemsep, topsep=0pt]
        \item In 2027, endogenous retirement enables emissions reductions nearly equal to the base case.
        \item This is almost entirely due to replacing approximately half of coal generation (mostly with gas generation).
        \item Emissions continue to drop through 2035, staying below the age-based retirement case.
        \item Nuclear capacity drops by more than half in 2040 once current policies stop preventing retirements.
        \item By 2045, the retirement case has higher emissions.
        \item Low gas prices lead to lower coal and nuclear capacity and slightly lower emissions in all periods.
        \item Raising coal prices by \$0.50/MMBTU retires all but 10 GW of coal capacity, replaced with natural gas and solar.
        %\item Slightly more transmission built.
    \end{itemize}
    \vspace{0.1em}
\end{minipage} \\ 
\raggedright Short Sample Period & 
\begin{minipage}[t]{\linewidth}
    \vspace{-\baselineskip}
    \begin{itemize}[leftmargin=*, noitemsep, topsep=0pt]
        \item Nationally, most models have less generation from wind and more from CCS when using fewer sample weeks.
        \item Emissions differ by less than 5\% in all model periods.
        \item Larger relative differences in capacity and generation decisions are seen in regional results.
    \end{itemize}
\end{minipage} & 
\begin{minipage}[t]{\linewidth}
    \vspace{-\baselineskip}
    \begin{itemize}[leftmargin=*, noitemsep, topsep=0pt]
        \item Nationally, all models have less wind and more solar generation when using fewer sample weeks.
        \item Emissions vary across models more than in the Net-Zero scenario.
        \item They also vary more from the 52-week configuration and are generally higher.
        \item Larger relative differences in capacity and generation decisions are seen in regional results.
    \end{itemize}
    \vspace{0.1em}
\end{minipage} \\ 
\raggedright Multi-Period Foresight & 
\begin{minipage}[t]{\linewidth}
    \vspace{-\baselineskip}
    \begin{itemize}[leftmargin=*, noitemsep, topsep=0pt]
        \item Small changes from myopic, but not consistent across models.
        \item GenX has less CCS generation, more from wind/solar.
        \item USENSYS is the opposite.
        \item GenX builds more transmission in foresight, Switch/USENSYS build less.
    \end{itemize}
\end{minipage} & 
\begin{minipage}[t]{\linewidth}
    \vspace{-\baselineskip}
    \begin{itemize}[leftmargin=*, noitemsep, topsep=0pt]
        \item TEMOA results suggest foresight leads to less wind/solar through 2035, then a large build-out in 2040 while plants still qualify for the PTC.
        \item GenX builds more wind and solar in early periods than TEMOA, but both show a similar build-out in 2040.
        \item The models disagree about whether emissions will be higher than the myopic configuration before 2040.
    \end{itemize}
\end{minipage} \\ 
\br
\end{tabular}
\end{table}

\section{Discussion}

The results of this intercomparison study provide significant insights for policymakers, researchers, and users of electricity CEMs interested in guiding the US power sector’s decarbonization. Our findings confirm that harmonizing model inputs and carefully differentiating between scenarios, configurations, and model structures can improve the interpretability and robustness of model outcomes. These practices help stakeholders make more confident and informed decisions in energy policy and system planning.

We summarize a few key findings from the more detailed descriptions in \Sref{section:results}. 

\begin{enumerate}
    \item Given the input cost assumptions, the models agree that a net-zero scenario is more expensive, has higher emissions, and relies more on inter-regional transmission when CCS is not available. Without CCS, the models select wind, solar, and battery storage as a least-cost substitute as opposed to nuclear power.
    \item Generators were assumed to retire either economically (when operating costs exceed revenue) or when they age out at the end of their useful service life, and these assumptions were the most influential configuration feature. Economic retirement reduces operational costs in both the current policies and net-zero scenarios. It also impacts investment decisions and emissions, especially in the current policies scenario. Allowing plants to retire economically immediately halves the existing coal fleet for current policies in the short run and nearly eliminates coal under the net-zero scenarios. Assuming current policies remain unchanged through 2050, however, economic retirement retains more coal capacity, in part to make up for nuclear plants that retire in 2040 when existing subsidies expire. Absent more aggressive decarbonization policies or additional economic pressures\footnote{see \ref{appendix:fuel_prices} for a discussion of how fuel prices affect coal, nuclear, and natural gas capacity with economic retirement}, the results imply that the economic stress on coal plants may be short-lived.
    \item The current policy scenario is sensitive to whether investment decisions are assumed to be made myopically or with foresight. With foresight, wind and solar expand less in the initial periods in anticipation of falling costs, then grow much faster in the final period where subsidies are available. The myopic configuration has a slower \textit{rate} of expansion, and ultimately builds less wind and solar capacity. However, foresight models are larger and require more computational resources because they account for investment and operational decisions across multiple periods, which can force modeling trade offs. Here, we model myopic and foresight configurations assuming 52-week and 20-week temporal resolutions, respectively. We find that the information gained with foresight does not necessarily compensate for the loss of temporal detail, with investment decisions from the 52-week myopic models often yielding lower system costs and less non-served energy than the 20-week foresight models.
    \item Preventing any expansion of inter-regional transmission (net-zero scenario only) increases 2050 operational costs by 4.6\% and emissions by 55\%. This implies that it is possible to substitute for transmission when decarbonizing, but at the expense of an ‘unmet policy penalty’. However, the total penalty appears to be modest. The assets that substitute for transmission vary regionally based upon resource adequacy, existing assets, and prices.
    \item The difference in total system costs between configurations is often less than the differences in investment decisions and emissions. 
\end{enumerate}

Harmonization has the important spillover benefit of distilling residual but important model differences, which deserve equal merit to the above consensus-based findings. After harmonization, the models still retained differences in the algebraic formulations of their native optimization routines. Some important differences include alternative dispatch configurations (the ability to simulate unit commitment), configurations of downtime, the co-optimization of storage power and capacity, and whether technologies vary by vintage or operational parameters vary by period. These residual differences were not sufficiently influential to preclude harmonizing our base configurations, but we would expect more differences between models if they were not forced into such strict alignment via data inputs. We emphasize that coding technology vintages and allowing operations to vary by period were particularly useful when parameterizing the complex subsidy designs of current policy production tax credits.

We reflect upon our harmonization effort to offer lessons learned for other groups engaged in similar future efforts. We found carefully differentiating between scenarios (representing different policy alternatives) from model configurations (alternative model resolutions and technical abstractions of the energy system) extremely helpful during project execution and for communicating results. The project also benefited from a shared data pipeline to ensure consistent inputs across models. The pipeline improved efficiency and led to a better understanding of the assumptions and methods of each model. There were considerable challenges in accommodating the various native model input formats with a data pipeline that was originally designed for one model (GenX).\footnote{Many of the difficulties with harmonizing data were related to translating operating parameters such as ramp rates, or labeling technologies as eligible for economic retirement.} Determining when differences in results were due to a missed or misinterpreted model parameter rather than underlying model differences---and finding the errant parameter---was an especially difficult task. If we had known that model results would align so well across nearly all configurations, we would have focused more on identifying input differences in data pipelines. Finally, the project benefited from a democratic and consensus-oriented decision-making process that enabled open discovery of the strengths, weaknesses, and differences of the models. This process yielded positive results, as all models improved---some even incorporating previously unconfigured features---and all modeling teams gained a deeper understanding of various approaches to representing real-world complexities within a mathematical framework.

We see tremendous value in the consensus-forming nature of model intercomparisons and harmonization. When research findings about specific energy or climate policies differ across studies, policy decisions regarding whether or how to implement these policies may lack robustness—particularly if differences remain unexplained or if policymakers poorly understand how strongly results depend on model configuration choices. Model intercomparisons, in which models use common input data and transparently document their configuration choices, offer one strategy to improve understanding and build consensus around robust results. Other strategies, such as establishing standards, including clear disclaimers, or forming decisions through committees, may also be valuable. As part of this effort, we developed a pipeline to render consistent input databases—a step toward establishing a common standard that aligns well with parallel efforts by others. Importantly, any standard-setting process should itself be consensus-oriented to ensure broad participation.

% \section{Acknowledgment}

% This work was supported by Grant Number G-2022-17196 from the Alfred P. Sloan Foundation. The authors gratefully acknowledge the Foundation’s support and commitment to advancing research in this field.

% \section{Data availability statement}
% Input data used by each of the models in this study are available at https://doi.org/10.5281/zenodo.14906951. Model results and dashboards are available at https://doi.org/10.5281/zenodo.14907274.

\appendix
% \section*{Appendix}
\section{Model regions}
\label{appendix:model_regions}
\begin{figure*}[ht]
\includegraphics[width=\textwidth,trim={0cm 0cm 0 0cm},clip]{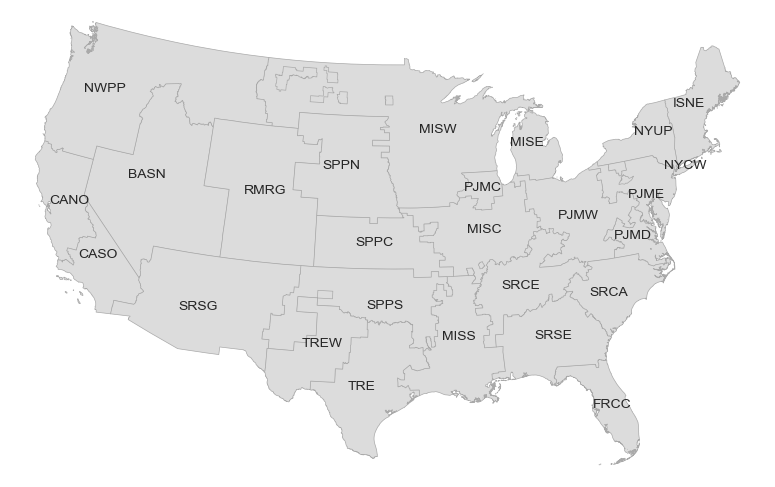}
\caption{Model regions.}
\label{fig:model-regions}
\end{figure*}

\begin{figure*}[ht]
\includegraphics[width=\textwidth,trim={0cm 0cm 0 0cm},clip]{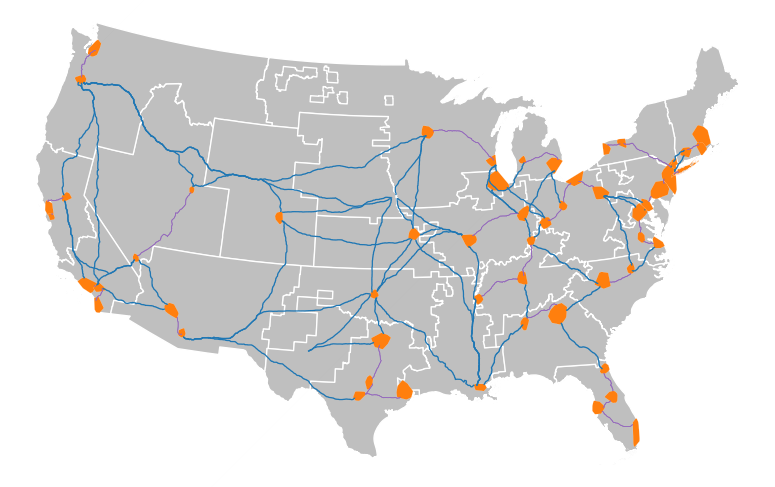}
\caption{Inter-regional (blue) and intra-regional (purple) transmission lines.}
\label{fig:model-transmission}
\end{figure*}

The model setup used here aggregates regions from EPA's Integrated Planning Model \cite{environmentalprotectionagencyIntegratedPlanningModel2003} into the 26 zones shown in \Fref{fig:model-regions}. We model transmission between zones transmission as a combination of inter-regional lines that connect large urban centers across regions and intra-regional transmission that connects large urban centers within regions (where applicable, see \Fref{fig:model-transmission}). Expansion costs and line losses account for 1 MW of intra-regional transmission per MW of inter-regional transmission. The starting transfer capacity, line losses, and cost of line reinforcement can be found at the project GitHub repository \cite{MIP_results_comparison}.

\section{Production tax credits}
\label{appendix:PTC}

As part of the Inflation Reduction Act, new zero-carbon resources are eligible for a production tax credit for the first 10 years of operation. Facilities that commence construction by 2036 are eligible for the PTC, so we include it for all new-build resources through 2040. Assuming that prevailing wage and apprenticeship requirements are met, the PTC value is \$27.50/MWh, with a 10\% bonus for projects located in an energy community.

Including the full PTC value in myopic would inflate its value since the PTC will only be in place for the first 10 years of operation. We account for this by calculating the present value of the credit (less a 7.5\% transfer penalty), and amortizing it over the 30 year facility lifetime. Technology WACCs from ATB 2022, which do not match the 5\% WACC used elsewhere in this study, were used for these calculations. The amortized PTC value is included as a negative variable cost for resources in model periods through 2040.

In foresight runs the PTC is nearly impossible to include if variable costs apply to resources built in every previous model period. Using the myopic model inputs, resources built in 2027 would get the credit for 16 years and resources built in 2040 would get the credit for 5 years. TEMOA and USENSYS were able to model the PTC for each vintage of wind and solar resources, providing the full credit for two model periods (10 years) and no credit after that point. Switch and GenX, which are not able to vintage new build resource, convert the PTC to an ITC.

\section{Net-zero CO$_2$ targets}
\label{appendix:co2_targets}

The net-zero CO$_2$ targets used in this study are based on results from the REPEAT study's Net-Zero Pathway \cite{jenkins_climate_2023}. REPEAT models emissions across the entire US energy system. The Net-Zero Pathway finds that the most cost-effective place to reduce emissions in the near term is in the power system. While REPEAT allows some emissions from the power sector in 2050, we modified the trajectory using a linear target to zero emissions from 2030 to 2050. Our 2027 target of 847 million tonnes CO$_2$ is erroneously based on the 2025 REPEAT value -- the results in 2027 are 494 million tonnes CO$_2$.

\begin{table}[!ht]
\caption{\label{appendix_co2_target}CO$_2$ targets by model year (Million Tonnes)}
    \footnotesize
    \begin{tabular}{@{}p{0.5in}p{0.75in}p{0.8in}}
    \br
    \textbf{Year} & \textbf{This study} & \textbf{REPEAT} \\ \hline
        2027 & 847 & 847 (in 2025) \\ 
        2030 & 186 & 186\\ 
        2035 & 130 & 130 \\ 
        2040 & 86.7 & 135 \\
        2045 & 43.3 & 113 \\
        2050 & 0 & 91 \\
    \br
    \end{tabular}
    % \end{indented}
\end{table}

\section{Varying fuel prices}
\label{appendix:fuel_prices}

When moving from age-based to endogenous retirement in the Current Policy scenario, our primary results show that more coal capacity is retained through 2050 and coal generation is higher when using endogenous retirement. Because the merit order dispatch of coal and gas plants can depend on fuel price, we tested alternate scenarios to show how changes in fuel price might affect the results. Low natural gas prices are based on AEO’s “High Oil and Gas Supply” case. High coal prices are based on adding \$0.50/MMBTU to the AEO reference coal price.

Lower gas prices lead to increased natural gas generation – but not more capacity – and more natural gas with CCS capacity/generation. This increase in generation from natural gas power plants displaces coal, nuclear, and wind from the system. It also reduces emissions in all study periods, with smaller reductions in 2045 and 2050.

\begin{figure*}[ht]
\includegraphics[scale=1,trim={0cm 0cm 0 0cm},clip]{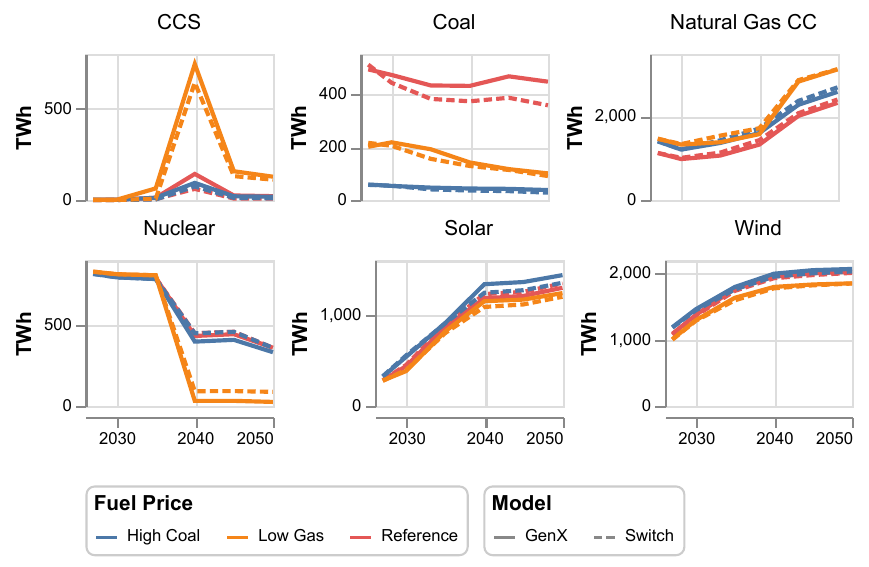}
\caption{Generation by source in the current policies scenario with economic retirement and alternative fuel prices.}
\label{fig:fuel-price-gen}
\end{figure*}

\begin{figure*}[ht]
\includegraphics[scale=1,trim={0cm 0cm 0 0cm},clip]{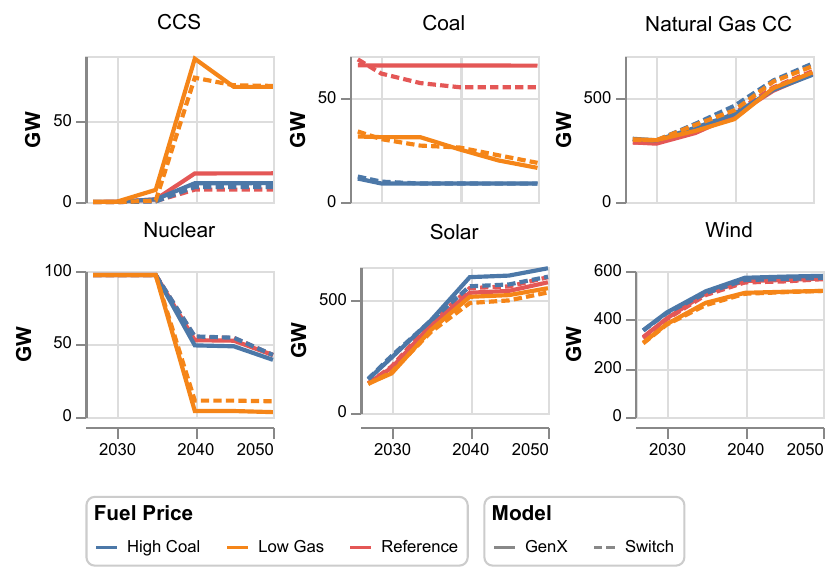}
\caption{Capacity by source in the current policies scenario with economic retirement and alternative fuel prices.}
\label{fig:fuel-price-cap}
\end{figure*}

\begin{figure*}[ht]
\includegraphics[scale=1,trim={0cm 0cm 0 0cm},clip]{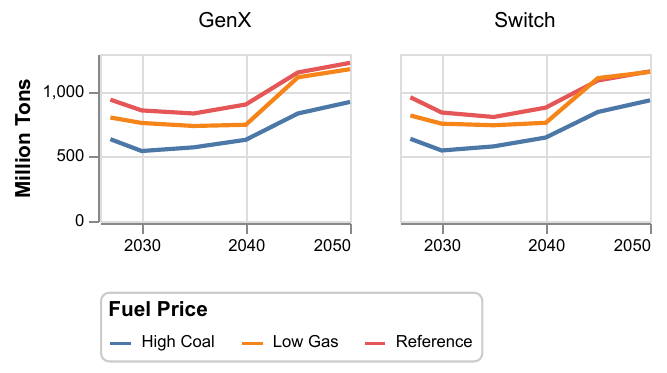}
\caption{Annual emissions within each planning period in the current policies scenario with economic retirement and alternative fuel prices.}
\label{fig:fuel-price-emiss}
\end{figure*}

\section{Harmonizing inputs across models}

Harmonizing inputs across diverse power system models revealed several key insights into how model design influences input requirements and configuration flexibility. Each model’s input structure is shaped by its development context and intended applications, leading to differences in how they handle planning period setups, resource retirements, and cost specifications. For instance, GenX, which typically operates in a myopic mode, organizes its inputs as separate files for each planning period. This setup simplifies modifying operational parameters, such as variable O\&M, across periods but complicates tracking capacity vintages within single resources. As a result, GenX often relies on economic retirement instead of age-based retirements, and power plant ages aren’t tracked, limiting certain analyses.

Switch, by contrast, is more commonly run in foresight mode and accommodates age-based retirements. Its input structure separates period-specific resource information from data constant across periods, facilitating long-term projections. While Switch can be adapted to a myopic setup by using multiple input files, this approach isn’t standard, which may affect ease of harmonization when comparing foresight and myopic models directly.

Temoa provides more extensive support for resource vintages and operating parameters, particularly useful for age-based retirements and variable cost changes over time, such as those influenced by production tax credits. With inputs stored in a single SQLite database, Temoa indexes parameters by both model period and technology vintage, though this level of detail can result in model “bloat.” Temoa’s foresight and optional limited myopic capabilities allow it to handle both economic and age-based retirements, and its ability to specify parameters such as WACC and O\&M costs by vintage aids in precision. However, unlike GenX, Temoa does not co-optimize power and energy capacities for storage, requiring users to specify different storage durations as separate resources.

USENSYS, built on the energyRt platform, provides substantial flexibility for model configuration, facilitating adaptation to project-specific assumptions. Its modular structure allows for a three-part representation of energy storage—"charger," "accumulator," and "discharger"—optimized independently for each storage capacity parameter. USENSYS applies an age-based framework for retirements, with options for early retirement, and can track capacities at various levels, treating vintages with unique parameters as distinct technologies. While designed for foresight applications, modifications in USENSYS support myopic optimization and time-sliced sub-annual modeling, expanding its flexibility but introducing complexity in harmonization with other models.

PowerGenome served as the primary data pipeline across models, yet harmonizing these inputs highlighted challenges, particularly where certain input data were relevant to only one model’s configuration. For example, GenX uses ramp rate limits only when unit commitment constraints are active, so this data might remain inactive in other configurations. A clearer distinction of which data inputs were essential to each configuration would have improved efficiency, reducing early misunderstandings and clarifying assumptions in the intercomparison process.

Ultimately, these differences underscore the importance of understanding model-specific input requirements when harmonizing for intermodel comparisons. For model users, documenting the relevance and function of each input field for each scenario and configuration can streamline cross-model integration and enhance interpretability across studies.

\newpage
% Note: \References (from iopart.cls) opens a harvard environment, but 
% entries from \bibliography{MIP-bib} are incompatible with this environment
% when using \bibliographystyle{iopart-num} (gives a "missing \item?"
% error). So we bypass it for now.
% \References
\section*{References}
\bibliography{MIP-bib}
% \References fails to close harvard environment (does \refs but not \endrefs)
% \end{harvard}
\end{document}